\documentclass[superscriptaddress,aps,prl,showpacs, amsmath,amssymb,
 reprint
]{revtex4-1}

\usepackage{graphicx}
\usepackage{dcolumn}
\usepackage{bm}

\begin{document}

\title{Spin Hall Magnetoimpedance}
\affiliation{Walther-Meissner-Institut, Bayerische Akademie der Wissenschaften, 85748 Garching, Germany}
\author{Johannes Lotze}
\affiliation{Walther-Meissner-Institut, Bayerische Akademie der Wissenschaften, 85748 Garching, Germany}
\author{Hans Huebl}
\affiliation{Walther-Meissner-Institut, Bayerische Akademie der Wissenschaften, 85748 Garching, Germany}
\affiliation{Nanosystems Initiative Munich (NIM), Schellingstra{\ss}e 4, 80799 M\"unchen, Germany}
\author{Rudolf Gross}
\affiliation{Walther-Meissner-Institut, Bayerische Akademie der Wissenschaften, 85748 Garching, Germany}
\affiliation{Nanosystems Initiative Munich (NIM), Schellingstra{\ss}e 4, 80799 M\"unchen, Germany}
\affiliation{Physik-Department, Technische Universit\"at M\"unchen, 85748 Garching, Germany}
\author{Sebastian T. B. Goennenwein}
\affiliation{Walther-Meissner-Institut, Bayerische Akademie der Wissenschaften, 85748 Garching, Germany}
\affiliation{Nanosystems Initiative Munich (NIM), Schellingstra{\ss}e 4, 80799 M\"unchen, Germany}
 \email{...}

\date{\today}

\begin{abstract}
The recently discovered spin Hall magnetoresistance effect electrically probes pure spin current flow across a ferrimagnetic insulator/normal metal bilayer interface. While usually the DC electrical resistance of the bilayer is measured as a function of the magnetization orientation in the magnetic insulator, we here present magnetoimpedance measurements using bias currents with frequencies up to several GHz. We find that the spin Hall magnetoresistance effect persists up to frequencies of at least 4\,GHz, enabling a fast readout of the magnetization direction in magnetic insulator/normal metal bilayers. Our data furthermore show that all interaction time constants relevant for the spin Hall magnetoresistance effect are shorter than 40\,ps.
\end{abstract}

\maketitle
The spin Hall effect (SHE) and its inverse describe the interconversion of charge and spin currents \cite{PhysRevLett.83.1834}. The SHE thus is of key importance for a broad variety of spin current-based and spin-caloritronic phenomena, such as the spin Seebeck effect \cite{apl/97/17/10.1063/1.3507386,PhysRevLett.108.106602,PhysRevB.88.094410,PhysRevB.88.064408,PhysRevB.87.054421,PhysRevLett.110.067206,PhysRevB.89.014416}, spin pumping \cite{kajiwaratransmission2010,PhysRevLett.104.046601, :/content/aip/journal/jap/109/7/10.1063/1.3549582,
PhysRevLett.111.247202,1401.6469,1307.2961,PhysRevLett.111.217204}, and spin Hall magnetoresistance \cite{PhysRevB.87.184421,PhysRevB.87.174417,PhysRevLett.110.206601,PhysRevB.87.224401,PhysRevB.87.144411}. Hereby, it is generally assumed that the spin Hall physics are independent of frequency up to tens or hundreds of GHz. Moreover, even optically detected voltages at THz frequencies \cite{kampfrathterahertz2013} have been interpreted in terms of the inverse SHE. Such a fast response of the spin Hall effect appears reasonable, since microscopic models attribute the SHE to spin-orbit coupling. However, an experimental investigation of this conjecture has not been put forward to our knowledge to date.

In order to critically test the presumed frequency-independence of spin Hall physics in the GHz frequency range, spin Hall magnetoresistance (SMR) experiments as a function of frequency appear particularly attractive. The SMR arises in ferromagnetic insulator/normal metal (FMI/N) \footnote{For the sake of simplicity, we use the term ``ferromagnetic'' in the sense of ``exhibiting long-range magnetic order''. In particular, we also refer to ferrimagnetic materials as ferromagnetic} bilayers. A pure spin current is sourced from the charge current flowing in the normal metal by the SHE. Depending on the orientation of the magnetization of the FMI with respect to the spin polarization of the spin current, this spin current can or cannot propagate across the interface into the insulating ferromagnetic layer. This results in a characteristic dependence of the resistance of the normal metal on the magnetization orientation in the adjacent magnetic insulator, although no electrical current flows through the FMI. Owing to this mechanism, the SMR exhibits a characteristic dependence on the magnetization orientation in the FMI, which is qualitatively different from anisotropic magnetoresistance in bulk polycrystalline FM metals \cite{PhysRevLett.110.206601,PhysRevB.87.224401,PhysRevB.87.174417,PhysRevB.87.184421}. Moreover, the SMR depends quadratically on the spin Hall angle $\theta_\mathrm{SH}$ \cite{PhysRevB.87.144411}, i.e. on spin Hall physics. The SMR thus is very sensitive to a possible change of the spin Hall effect viz. the spin Hall angle $\theta_\mathrm{SH}$ with frequency. Furthermore, experiments as a function of the AC current frequency allow for  testing of the viability of the SHE for high-frequency all-electrical spin current generation, and for SMR-based fast readout of the magnetization orientation of an insulating ferromagnet, which is desirable for use in spintronic devices.

In this work, we perform magnetoimpedance measurements by applying an AC charge current with frequency $f$ to a yttrium iron garnet/platinum (YIG/Pt) hybrid bilayer, and investigate how the resistance $R(f,\mathbf{M})$ of the hybrid changes both as a function of frequency, and as a function of the orientation of the magnetization $\mathbf{M}$ in the YIG film. Our data, recorded at room temperature, invariably exhibit the evolution of the resistance with magnetization orientation characteristic of SMR for charge current frequencies from DC to 4\,GHz. In other words, the magneto-resistive response of our YIG/Pt hybrid (viz. the SMR effect) does not depend on frequency to within experimental accuracy up to frequencies of at least 4\,GHz.

\begin{figure}
\includegraphics{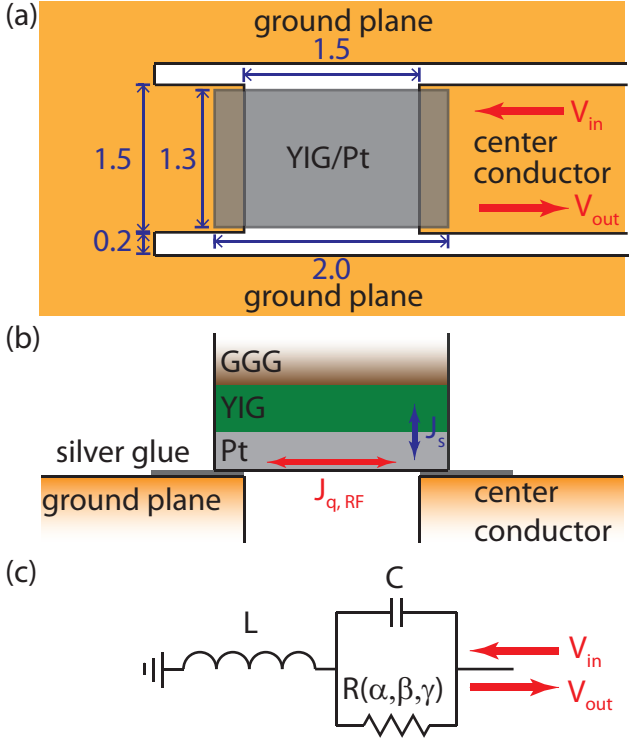}
\caption{\label{fig:setup} (color online) (a,b) YIG/Pt bilayer bridging a gap in the Cu center conductor of a coplanar waveguide (CPW) structure shown in (a) top view, (b) cross-sectional drawing along the center conductor. The dimensions shown are in mm.(c) The equivalent electrical circuit model used to describe the YIG/Pt bilayer on the CPW.}
\end{figure}

The YIG/Pt bilayer was prepared by growing a $55\,\mathrm{nm}$ thick YIG film on a (111)-oriented single-crystalline gadolinium gallium garnet substrate using laser-MBE \cite{0022-3727-45-3-033001,PhysRevB.87.224401}. Without breaking the vacuum, the sample was subsequently transferred to an electron beam evaporation chamber and a Pt film with a thickness of $4\,\mathrm{nm}$ was deposited onto the YIG. More details on the sample preparation procedure are given in Ref.~\cite{PhysRevB.87.224401}. For the experiments discussed here, we diced a rectangular piece with lateral dimensions of $2 \times 1.3 \,\mathrm{mm^2}$ from the as grown sample.

To measure the impedance of this YIG/Pt sample up to GHz frequencies, we integrate it into a coplanar waveguide (CPW) structure. The latter has a characteristic impedance of $50\,\mathrm{\Omega}$ and thus allows for the propagation of a high frequency charge current in a broad frequency range. The CPW structure was patterned onto a printed circuit board (PCB) as shown in Fig.~\ref{fig:setup}(a,b). The width of the center conductor of the CPW is $1.5\,\mathrm{mm}$, with a gap of $0.2\,\mathrm{mm}$ to the ground planes. The AC current is injected using a surface mount mini-SMP connector at one end of the CPW structure. The CPW is short-circuited at the other end. The center conductor is interrupted by a $1.5\times 1.5\,\mathrm{mm^2}$ square gap in the PCB board. The YIG/Pt bilayer is attached to the CPW structure with the Pt facing down toward the copper of the CPW centerline using silver glue, bridging this gap, as shown in Fig.~\ref{fig:setup}(a,b). Since the sample that is integrated in this way into the CPW structure constitutes a load that is not equal to the system impedance $Z_0=50\,\mathrm{\Omega}$, part of the AC current is reflected at the sample. Measuring this reflection allows us to extract the impedance of the sample.

\begin{figure*}
\includegraphics[scale=1]{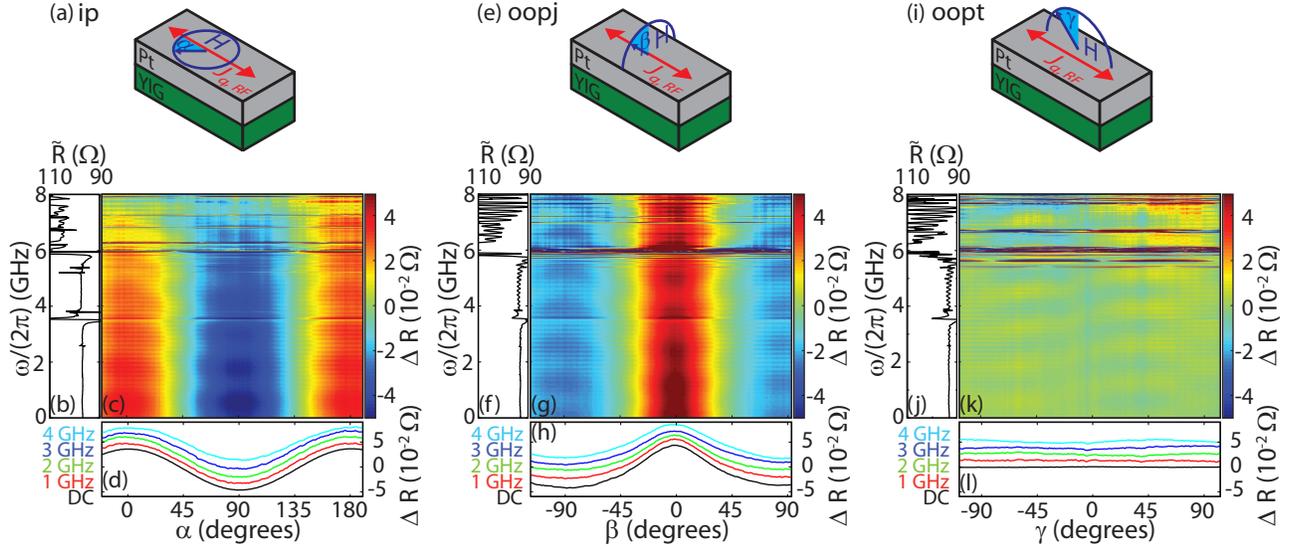}
\caption{\label{fig:all} (color online) Frequency dependent resistance for the ip, oopj, and oopt rotation planes. Panels (a,e,i) show a sketch of the YIG/Pt bilayer and the external magnetic field relative to the applied bias current direction. $\alpha$ parametrizes the angle of the magnetic field in the ip rotation, $\beta$ in the oopj rotation, and $\gamma$ in the oopt rotation. Panels (b,f,j) show $\widetilde{R}$ (Eq.~(\ref{eq:tilR})) from frequencies of DC to 8\,GHz for the respective magnetic field rotations. Panels (c,g,k) show the resistance modulation $\Delta R$ with respect to AC current frequency and the corresponding magnetic field rotation angles at a constant external magnetic field of $\mu_0 |\mathbf{H}|= 0.6\,\mathrm{T}$. Panels (d,h,l) show $\Delta R$ as a function of the respective rotation angles at different, fixed frequencies: DC (black line), $1\,\mathrm{GHz}$ (red line), $2\, \mathrm{GHz}$ (green line), $3\, \mathrm{GHz}$ (blue line), $4\, \mathrm{GHz}$ (light blue line). The $\Delta R$ curves are offset for clarity.}
\end{figure*}

The CPW/sample chip is then inserted into the magnetic field of a rotatable electromagnet. We mounted the sample chip in three different ways (see Figs.~\ref{fig:all}(a, e, i)): In the in-plane (ip) configuration, the magnet rotation axis is parallel to the film normal, so that the magnetic field is always in the plane of the YIG/Pt bilayer. The magnetic field direction is parametrized by the angle $\alpha$ between the charge current direction and the magnetic field direction. In the oopj configuration the rotation axis of the magnetic field is in the plane, parallel to the current direction, with the angle $\beta$ between the magnetic field and the film normal. This results in a magnetic field that is rotated from an out of plane orientation to an in plane orientation relative to the bilayer. In the oopt configuration, the rotation axis lies in the film plane, perpendicular to the current direction. The magnetization direction is represented by the angle $\gamma$ between the magnetic field direction and the film normal.
For a (DC) SMR-like behavior \cite{PhysRevB.87.224401,PhysRevB.87.174417}, we expect a $\cos^2(\alpha)$-like resistance modulation with amplitude $R_1$ on a constant offset $R_0$ upon rotating the magnetization in the film plane:
\begin{equation}
R_\mathrm{ip}(\alpha)=R_0+R_1\cos^2\alpha\, .
\label{eq:ip}
\end{equation}
The ratio
\begin{equation}
\frac{R_1}{R_0}=\frac{2 \theta_\mathrm{SH}^2 \lambda_\mathrm{SD}^2 \rho t^{-1}G_\mathrm{r} \tanh^2(\frac{t}{2 \lambda_\mathrm{SD}})}{1+ 2\lambda_\mathrm{SD}\rho G_\mathrm{r} \coth (\frac{t}{\lambda_\mathrm{SD}})}
\label{eq:MRrat}
\end{equation}
depends on the spin Hall angle $\theta_\mathrm{SH}$, the resistivity $\rho$ of the Pt, the spin diffusion length $\lambda_\mathrm{SD}$, the real part of the spin mixing interface conductance $G_\mathrm{r}$ \cite{PhysRevB.87.144411}, and the thickness $t$ of the Pt film. As usually done in the literature \cite{PhysRevLett.110.206601,PhysRevB.87.224401,PhysRevB.87.174417,PhysRevB.87.184421}, we here take all of these parameters as constants, independent of frequency and magnetic field. We furthermore assume that $\theta_\mathrm{SH}$ is purely real. For the oopj rotation, we expect
\begin{equation}
R_\mathrm{oopj}(\beta)=R_0+R_1\cos^2\beta\, ,
\label{eq:oopj}
\end{equation}
where $\beta$ defined as sketched in Fig.~\ref{fig:all}(e). In the oopt rotation, the SMR is independent of the magnetization orientation \cite{PhysRevB.87.144411} with
\begin{equation}
R_\mathrm{oopt}=R_0+R_1\,.
\label{eq:oopt}
\end{equation}

To establish a reference for the AC resistance measurements, we first measured the DC resistance as a function of the magnetic field orientation at a fixed magnetic field magnitude $\mu_0 H=0.6\, \mathrm{T}$ for all three magnetic field rotation configurations. In these experiments, a constant bias charge current is applied to the CPW strip with a Keithley 2400 sourcemeter, and the resistance is calculated from the voltage drop.

In a second set of experiments, we measured the complex reflection coefficient $S_{11}$ with an Agilent N5242A vector network analyzer (VNA) as a function of frequency $f=\omega/(2\pi)$, and as a function of the magnetic field orientation angles $\alpha,\,\beta$, and $\gamma$. Again, the magnetic field magnitude hereby was $0.6\,\mathrm{T}$. More precisely, for each measured magnetic field orientation, the frequency of the VNA microwave drive signal is swept, and the corresponding $S_{11}(\omega)$ recorded. Then the magnetic field is rotated to the next orientation, $S_{11}(\omega)$ is recorded, etc. We calibrate the RF circuitry using a set of homemade calibration standards consisting of a CPW structure identical to the one on which the sample is mounted, but without a sample on it (open), another with bond wires connecting the center conductor to the ground plane directly at the interruption of the center conductor (short), as well as one with two $100\,\Omega$ high-frequency resistors bridging the gap to ground on either side before the gap in the center conductor, resulting in a $50\,\Omega$ load. Reference measurements with these calibration standards allow us to calibrate the signal path up to the sample position. Ideally, after subtraction of the calibration data, the $S_{11}$ measured with the sample chip then only reflects the properties of the YIG/Pt sample.

The measured complex scattering parameter $S_{11}$ (corrected using the calibration data) is converted to the complex impedance $Z$ of the sample via \cite{pozar2011microwave}
\begin{equation}
Z(\omega)=\frac{Z_0(1+S_{11}(\omega))}{1-S_{11}(\omega)}\, , 
\label{eq:ZfromS}
\end{equation}
where $Z_0=50\,\Omega$ is the characteristic impedance of the system. The measurement of the impedance of the sample via the $S_{11}$ scattering parameter has two advantages: firstly, compared to a transmission measurement the slope of 
\begin{equation}
S_{11}(Z)=\frac{Z-Z_0}{Z+Z_0}
\end{equation} 
is steeper near the characteristic impedance $Z_0$, leading to a higher measurement sensitivity. Secondly, this measurement is easier to calibrate, requiring only one set of open, short and load calibration measurements and eliminating the need of a through calibration measurement \cite{dunsmore2012handbook}.
To extract the magnetization orientation dependent resistance $R$ of the YIG/Pt bilayer from the complex impedance, we use the circuit model sketched in Fig.~\ref{fig:setup}(c). $L$ and $C$ hereby are an inductance and a capacitance, respectively, taken as frequency independent constants. This model is consistent with models applied to surface mount resistors \cite{sengupta2005applied}. The impedance of this $L$-$R$-$C$ circuit shown in Fig.~\ref{fig:setup}(c) is given by 
\begin{equation}
Z(\omega)=\frac{1}{R \left(C^2 \omega^2+\frac{1}{R^2}\right)}+i \left(\omega L-\frac{C \omega}{C^2 \omega^2+\frac{1}{R^2}}\right)\,.
\label{eq:fit}
\end{equation}
In a first step, we calculate $Z(\omega)$ from the measurement data via Eq.~(\ref{eq:ZfromS}). We then simultaneously fit $\Re (Z)$ and $\Im (Z)$ with Eq.~(\ref{eq:fit}), using $R$, $L$, and $C$ as fit parameters. Since at higher frequencies resonance phenomena occur, which cannot be reproduced by the equivalent circuit model, only the part of $Z(\omega)$ with $\omega/(2\pi)<3\,\mathrm{GHz}$ is included in the fit.
\begin{figure}
\includegraphics{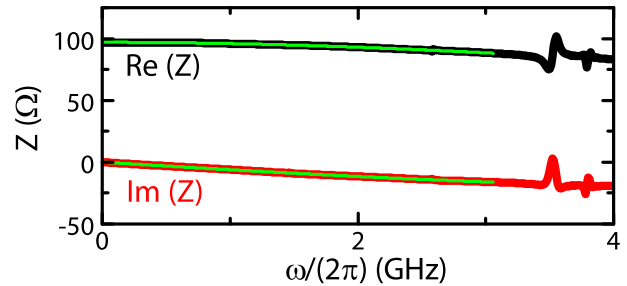}
\caption{\label{fig:fit} (color online) The real and imaginary parts of the complex impedance $Z(\omega)$ recorded for $\mu_0 |\mathbf{H}|=0.6\,\mathrm{T}$ and $\alpha=-90^\circ$ in the ip rotation measurement. Both are fitted simultaneously with Eq.~(\ref{eq:fit}) (green line), yielding the capacitance $C=0.2\,\mathrm{pF}$ and inductance $L=1\, \mathrm{nH}$ of the equivalent circuit of Fig.~\ref{fig:setup}(c).}
\end{figure}
This is exemplarily illustrated in Fig.~\ref{fig:fit}, where the real and imaginary parts of the complex impedance as well as the fit according to Eq.~(\ref{eq:fit}) (green line) are plotted for the measurement with the external field in the film plane. For both the oopj and oopt rotation (i.e., the rotations of the magnetic field from in- to out-of-plane) the data and fit look very similar. 
For all 3 rotation planes we find that $C=2\times 10^{-13}\,\mathrm{F}$ and $L=1\times 10^{-9}\,\mathrm{H}$ consistently describe the data. We did not find $L$ or $C$ to be magnetization orientation dependent. The parameter $R$ is found from the fit to be $R=97\,\Omega$, corresponding to the measured DC resistance of the device.
 
The total resistance $R$ consists of two components: the resistance $R_0$ of the Pt film which is independent of frequency and magnetic field, and $R_1$, which is magnetization orientation dependent. The (possible) frequency dependence of $R_1$ is the key focus of this paper. $R_1$ can be taken as small compared to $R_0$, because the magnetoresistance ratios $R_1/R_0$ measured in YIG/Pt are smaller than $10^{-2}$ \cite{PhysRevB.87.184421,PhysRevB.87.174417,PhysRevLett.110.206601,PhysRevB.87.224401}. Using $L$ and $C$, we can in a second step calculate the magnetization orientation dependent resistance from the measured impedance by solving Eq.~(\ref{eq:fit}) for $R$:
\begin{align}
&R(\omega,\{\alpha,\beta,\gamma\})=& \nonumber \\
&\frac{\sqrt{ L^2 \omega^2-|Z(\omega,\{\alpha,\beta,\gamma\})|^2}}{\sqrt{ C^2 \omega^2 \left(|Z(\omega,\{\alpha,\beta,\gamma\})|^2- L^2 \omega^2
  \right)+2 L C \omega^2 -1}}\, .
\label{eq:res}
\end{align} 
$R(\omega,\{\alpha,\beta,\gamma\})$ includes the frequency and magnetic field independent DC resistance $R_0$ of the platinum and the magnetization orientation dependent resistance $R_1$. 

In Fig.~\ref{fig:all}, we show the AC magnetoresistance obtained from our measurements for the three rotation planes. This figure is organized as follows: there are 4 panels for each rotation plane, respectively: in (a,e,i) we show a sketch of the measurement geometry for the three rotation planes and the rotation angle of the external magnetic field. In (b,f,j) we show the frequency dependent resistance averaged over all $N$ magnetization orientations studied in a given magnetic field rotation plane
\begin{equation}
\widetilde{R}(\omega)=\frac{1}{N}\sum_{i=1}^{N}R(\omega,\{\alpha_i,\beta_i,\gamma_i\})
\label{eq:tilR}
\end{equation} 
as a function of AC current frequency.

Panels (c,g,k) show the resistance modulation $\Delta R$, 
\begin{equation}
\Delta R(\omega,\{\alpha,\beta,\gamma\})=R(\omega,\{\alpha,\beta,\gamma\})-\widetilde{R}(\omega)\, ,
\label{eq:DR}
\end{equation}
as a function of both frequency and magnetic field angle in a false color plot, while $\Delta R$ traces recorded at selected frequencies are depicted in panels (d,h,l).

Due to calibration issues for $\omega/(2\pi)>3\,\mathrm{GHz}$, the resistance calculated from Eq.~(\ref{eq:res}) diverges at certain frequencies corresponding to standing wave resonances arising from reflections at the interface of the CPW structure and the sample. These background oscillations are suppressed in $\Delta R$, which makes it possible to plot $\Delta R$ in the false color plots of Figs.~\ref{fig:all}(c,g,k) using the same color code over the whole frequency range used in experiment.

We first analyze the change in DC resistance as a function of the magnetization orientation. The black lines in Figs.~\ref{fig:all}(d,h,l) show the change in resistance $\Delta R(DC)$ for $\mu_0 |\mathbf{H}|=0.6 \,\mathrm{T}$ as a function of the angle. The characteristic $\cos^2(\alpha)$-dependence of Eq.~(\ref{eq:ip}) is clearly evident in Fig.~\ref{fig:all}(d) as well as the expected $\cos^2(\beta)$ type modulation of Eq.~(\ref{eq:oopj}) for the oopj rotation in Fig.~\ref{fig:all}(h). In the latter case, the $\cos^2(\beta)$ modulation is not ideal, due to shape anisotropy, which prevents the magnetization from fully aligning with the applied magnetic field when it is not in the sample plane ($\beta=0^{\circ}$ corresponds to $\mathbf{H}$ along the film normal). Last but not least, for a rotation of the $0.6\,\mathrm{T}$ magnetic field in the oopt rotation plane, 
the resistance is constant (Fig.~\ref{fig:all}(l)), as expected from Eq.~(\ref{eq:oopt}). Thus the observed angular dependence is the one expected from the SMR effect according to Eqs.~(\ref{eq:ip})-(\ref{eq:oopt}). The DC resistance $R_0=97\,\mathrm{\Omega}$ and a resistance modulation amplitude $R_1=max(\Delta R)-min(\Delta R)=0.083\,\mathrm{\Omega}$ yield a MR ratio of $R_1/R_0=8.6 \times 10^{-4}$. Using the parameters $\theta_\mathrm{SH}=0.11$, $\lambda_\mathrm{SD}=1.5\,\mathrm{nm}$, $G_\mathrm{r}=4\times 10^{14}\,\mathrm{\Omega}^{-1}\, \mathrm{m^2}$ \cite{PhysRevB.87.224401,PhysRevLett.111.176601} and the thickness of the Pt film of $t=4\,\mathrm{nm}$, one expects a DC SMR magnitude of $R_1/R_0=7.7\times 10^{-4}$ from Eq.~(\ref{eq:MRrat}), in good agreement to the MR ratio measured experimentally.

We find that the phenomenology of the magnetoresistance observed does not change when making the transition from DC to AC bias currents. In the ip rotation of the external magnetic field, shown in Fig.~\ref{fig:all}(c), we find a modulation of the resistance with a $\cos^2(\alpha)$ dependency, regardless of the AC current frequency, up to at least 4\,GHz. Similarly, the oopj data (Fig.~\ref{fig:all}(g)) show a $\cos^2(\beta)$ dependency, while in the oopt orientation (Fig.~\ref{fig:all}(k)) the resistance is independent of magnetization orientation. In Figs.~\ref{fig:all}(d,h,l), we compare the change in resistance with respect to the applied magnetic field angle at DC as well as 1\,GHz, 2\,GHz, 3\,GHz, and 4\,GHz AC currents: the curve shape and the amplitude of the modulation is the same, irrespective of frequency.
Qualitatively, this modulation persists at frequencies higher than 4\,GHz. However, due to the homemade calibration standards and CPW structures, the resistance extraction becomes increasingly unreliable above 4\,GHz. We therefore attribute the slight decrease of the resistance modulation amplitude at higher frequencies to imperfect calibration or circuit design issues, rather than a frequency dependence of the mechanisms governing the SMR effect. We thus find that the phenomenology of the SMR can be described up to frequencies of at least 4\,GHz with real, frequency independent values for $L$, $C$, $R$, as well as $\theta_\mathrm{SH}$, of which only the resistance $R$ is magnetization orientation dependent.

The fact that the SMR effect persists up to at least $\omega/(2\pi)=4\,\mathrm{GHz}$ means that the interaction time constants $\tau=1/\omega$ relevant for the SMR effect are shorter than 40\,ps. Since the SMR requires both the spin Hall effect and spin torque transfer \cite{PhysRevB.87.144411,PhysRevB.87.174411,PhysRevB.88.214417}, i.e. spin-orbit interaction, this can be compared with the spin-orbit interaction time $\tau_\mathrm{SO}$ in platinum. In the free electron model $(2 \pi \tau_\mathrm{SO})^{-1}$ is estimated to be in the tens of GHz \cite{sakurai1994modern}, and much shorter $\tau_\mathrm{SO}$ are inferred from spin injection viz. spin transport experiments \cite{1468-6996-9-1-014105}. A constant SMR magnitude up to tens of GHz thus appears reasonable.

From a more applied perspective, our experiments show that the SMR can be used to read out the orientation in a ferromagnetic insulator such as YIG electrically in no more than 40\,ps.

In summary, we have measured the SMR effect in a YIG/Pt bilayer, using currents with frequencies from DC up to 8\,GHz. We can describe our results with a simple $L$-$R$-$C$ circuit model with frequency independent constants, of which only the resistance $R$ is magnetization dependent. We find a SMR amplitude (magnetoresistance ratio) of $8.6 \times 10^{-4}$, which is unaltered from DC up to frequencies of several GHz. This implies that the spin Hall physics enabling the SMR effect are frequency independent up to frequencies of at least 4\,GHz. This is consistent with theoretical work proposing that the time constants of the SMR should be governed by the spin-orbit interaction. Furthermore, our experiments demonstrate that AC SMR experiments enable an electrical read out of the magnetization orientation in magnetic insulators on timescales of a few 10\,ps or even less. 

We thank Stephan Gepr\"ags and Stephan Altmannshofer for the fabrication of the YIG/Pt bilayer. Financial support by Deutsche Forschungsgemeinschaft via SPP 1538 "Spin-Caloric Transport" (Project  No. GO 944/4-1) is gratefully acknowledged.

%


\begin{thebibliography}{31}%
\makeatletter
\providecommand \@ifxundefined [1]{%
 \@ifx{#1\undefined}
}%
\providecommand \@ifnum [1]{%
 \ifnum #1\expandafter \@firstoftwo
 \else \expandafter \@secondoftwo
 \fi
}%
\providecommand \@ifx [1]{%
 \ifx #1\expandafter \@firstoftwo
 \else \expandafter \@secondoftwo
 \fi
}%
\providecommand \natexlab [1]{#1}%
\providecommand \enquote  [1]{``#1''}%
\providecommand \bibnamefont  [1]{#1}%
\providecommand \bibfnamefont [1]{#1}%
\providecommand \citenamefont [1]{#1}%
\providecommand \href@noop [0]{\@secondoftwo}%
\providecommand \href [0]{\begingroup \@sanitize@url \@href}%
\providecommand \@href[1]{\@@startlink{#1}\@@href}%
\providecommand \@@href[1]{\endgroup#1\@@endlink}%
\providecommand \@sanitize@url [0]{\catcode `\\12\catcode `\$12\catcode
  `\&12\catcode `\#12\catcode `\^12\catcode `\_12\catcode `\%12\relax}%
\providecommand \@@startlink[1]{}%
\providecommand \@@endlink[0]{}%
\providecommand \url  [0]{\begingroup\@sanitize@url \@url }%
\providecommand \@url [1]{\endgroup\@href {#1}{\urlprefix }}%
\providecommand \urlprefix  [0]{URL }%
\providecommand \Eprint [0]{\href }%
\providecommand \doibase [0]{http://dx.doi.org/}%
\providecommand \selectlanguage [0]{\@gobble}%
\providecommand \bibinfo  [0]{\@secondoftwo}%
\providecommand \bibfield  [0]{\@secondoftwo}%
\providecommand \translation [1]{[#1]}%
\providecommand \BibitemOpen [0]{}%
\providecommand \bibitemStop [0]{}%
\providecommand \bibitemNoStop [0]{.\EOS\space}%
\providecommand \EOS [0]{\spacefactor3000\relax}%
\providecommand \BibitemShut  [1]{\csname bibitem#1\endcsname}%
\let\auto@bib@innerbib\@empty
\bibitem [{\citenamefont {Hirsch}(1999)}]{PhysRevLett.83.1834}%
  \BibitemOpen
  \bibfield  {author} {\bibinfo {author} {\bibfnamefont {J.~E.}\ \bibnamefont
  {Hirsch}},\ }\href {\doibase 10.1103/PhysRevLett.83.1834} {\bibfield
  {journal} {\bibinfo  {journal} {Phys. Rev. Lett.}\ }\textbf {\bibinfo
  {volume} {83}},\ \bibinfo {pages} {1834} (\bibinfo {year}
  {1999})}\BibitemShut {NoStop}%
\bibitem [{\citenamefont {Uchida}\ \emph {et~al.}(2010)\citenamefont {Uchida},
  \citenamefont {Adachi}, \citenamefont {Ota}, \citenamefont {Nakayama},
  \citenamefont {Maekawa},\ and\ \citenamefont
  {Saitoh}}]{apl/97/17/10.1063/1.3507386}%
  \BibitemOpen
  \bibfield  {author} {\bibinfo {author} {\bibfnamefont {K.}~\bibnamefont
  {Uchida}}, \bibinfo {author} {\bibfnamefont {H.}~\bibnamefont {Adachi}},
  \bibinfo {author} {\bibfnamefont {T.}~\bibnamefont {Ota}}, \bibinfo {author}
  {\bibfnamefont {H.}~\bibnamefont {Nakayama}}, \bibinfo {author}
  {\bibfnamefont {S.}~\bibnamefont {Maekawa}}, \ and\ \bibinfo {author}
  {\bibfnamefont {E.}~\bibnamefont {Saitoh}},\ }\href {\doibase
  http://dx.doi.org/10.1063/1.3507386} {\bibfield  {journal} {\bibinfo
  {journal} {Appl. Phys. Lett.}\ }\textbf {\bibinfo {volume} {97}},\ \bibinfo
  {eid} {172505} (\bibinfo {year} {2010})}\BibitemShut {NoStop}%
\bibitem [{\citenamefont {Weiler}\ \emph {et~al.}(2012)\citenamefont {Weiler},
  \citenamefont {Althammer}, \citenamefont {Czeschka}, \citenamefont {Huebl},
  \citenamefont {Wagner}, \citenamefont {Opel}, \citenamefont {Imort},
  \citenamefont {Reiss}, \citenamefont {Thomas}, \citenamefont {Gross},\ and\
  \citenamefont {Goennenwein}}]{PhysRevLett.108.106602}%
  \BibitemOpen
  \bibfield  {author} {\bibinfo {author} {\bibfnamefont {M.}~\bibnamefont
  {Weiler}}, \bibinfo {author} {\bibfnamefont {M.}~\bibnamefont {Althammer}},
  \bibinfo {author} {\bibfnamefont {F.~D.}\ \bibnamefont {Czeschka}}, \bibinfo
  {author} {\bibfnamefont {H.}~\bibnamefont {Huebl}}, \bibinfo {author}
  {\bibfnamefont {M.~S.}\ \bibnamefont {Wagner}}, \bibinfo {author}
  {\bibfnamefont {M.}~\bibnamefont {Opel}}, \bibinfo {author} {\bibfnamefont
  {I.-M.}\ \bibnamefont {Imort}}, \bibinfo {author} {\bibfnamefont
  {G.}~\bibnamefont {Reiss}}, \bibinfo {author} {\bibfnamefont
  {A.}~\bibnamefont {Thomas}}, \bibinfo {author} {\bibfnamefont
  {R.}~\bibnamefont {Gross}}, \ and\ \bibinfo {author} {\bibfnamefont
  {S.~T.~B.}\ \bibnamefont {Goennenwein}},\ }\href {\doibase
  10.1103/PhysRevLett.108.106602} {\bibfield  {journal} {\bibinfo  {journal}
  {Phys. Rev. Lett.}\ }\textbf {\bibinfo {volume} {108}},\ \bibinfo {pages}
  {106602} (\bibinfo {year} {2012})}\BibitemShut {NoStop}%
\bibitem [{\citenamefont {Schreier}\ \emph {et~al.}(2013)\citenamefont
  {Schreier}, \citenamefont {Kamra}, \citenamefont {Weiler}, \citenamefont
  {Xiao}, \citenamefont {Bauer}, \citenamefont {Gross},\ and\ \citenamefont
  {Goennenwein}}]{PhysRevB.88.094410}%
  \BibitemOpen
  \bibfield  {author} {\bibinfo {author} {\bibfnamefont {M.}~\bibnamefont
  {Schreier}}, \bibinfo {author} {\bibfnamefont {A.}~\bibnamefont {Kamra}},
  \bibinfo {author} {\bibfnamefont {M.}~\bibnamefont {Weiler}}, \bibinfo
  {author} {\bibfnamefont {J.}~\bibnamefont {Xiao}}, \bibinfo {author}
  {\bibfnamefont {G.~E.~W.}\ \bibnamefont {Bauer}}, \bibinfo {author}
  {\bibfnamefont {R.}~\bibnamefont {Gross}}, \ and\ \bibinfo {author}
  {\bibfnamefont {S.~T.~B.}\ \bibnamefont {Goennenwein}},\ }\href {\doibase
  10.1103/PhysRevB.88.094410} {\bibfield  {journal} {\bibinfo  {journal} {Phys.
  Rev. B}\ }\textbf {\bibinfo {volume} {88}},\ \bibinfo {pages} {094410}
  (\bibinfo {year} {2013})}\BibitemShut {NoStop}%
\bibitem [{\citenamefont {Hoffman}\ \emph {et~al.}(2013)\citenamefont
  {Hoffman}, \citenamefont {Sato},\ and\ \citenamefont
  {Tserkovnyak}}]{PhysRevB.88.064408}%
  \BibitemOpen
  \bibfield  {author} {\bibinfo {author} {\bibfnamefont {S.}~\bibnamefont
  {Hoffman}}, \bibinfo {author} {\bibfnamefont {K.}~\bibnamefont {Sato}}, \
  and\ \bibinfo {author} {\bibfnamefont {Y.}~\bibnamefont {Tserkovnyak}},\
  }\href {\doibase 10.1103/PhysRevB.88.064408} {\bibfield  {journal} {\bibinfo
  {journal} {Phys. Rev. B}\ }\textbf {\bibinfo {volume} {88}},\ \bibinfo
  {pages} {064408} (\bibinfo {year} {2013})}\BibitemShut {NoStop}%
\bibitem [{\citenamefont {Meier}\ \emph {et~al.}(2013)\citenamefont {Meier},
  \citenamefont {Kuschel}, \citenamefont {Shen}, \citenamefont {Gupta},
  \citenamefont {Kikkawa}, \citenamefont {Uchida}, \citenamefont {Saitoh},
  \citenamefont {Schmalhorst},\ and\ \citenamefont
  {Reiss}}]{PhysRevB.87.054421}%
  \BibitemOpen
  \bibfield  {author} {\bibinfo {author} {\bibfnamefont {D.}~\bibnamefont
  {Meier}}, \bibinfo {author} {\bibfnamefont {T.}~\bibnamefont {Kuschel}},
  \bibinfo {author} {\bibfnamefont {L.}~\bibnamefont {Shen}}, \bibinfo {author}
  {\bibfnamefont {A.}~\bibnamefont {Gupta}}, \bibinfo {author} {\bibfnamefont
  {T.}~\bibnamefont {Kikkawa}}, \bibinfo {author} {\bibfnamefont
  {K.}~\bibnamefont {Uchida}}, \bibinfo {author} {\bibfnamefont
  {E.}~\bibnamefont {Saitoh}}, \bibinfo {author} {\bibfnamefont {J.-M.}\
  \bibnamefont {Schmalhorst}}, \ and\ \bibinfo {author} {\bibfnamefont
  {G.}~\bibnamefont {Reiss}},\ }\href {\doibase 10.1103/PhysRevB.87.054421}
  {\bibfield  {journal} {\bibinfo  {journal} {Phys. Rev. B}\ }\textbf {\bibinfo
  {volume} {87}},\ \bibinfo {pages} {054421} (\bibinfo {year}
  {2013})}\BibitemShut {NoStop}%
\bibitem [{\citenamefont {Qu}\ \emph {et~al.}(2013)\citenamefont {Qu},
  \citenamefont {Huang}, \citenamefont {Hu}, \citenamefont {Wu},\ and\
  \citenamefont {Chien}}]{PhysRevLett.110.067206}%
  \BibitemOpen
  \bibfield  {author} {\bibinfo {author} {\bibfnamefont {D.}~\bibnamefont
  {Qu}}, \bibinfo {author} {\bibfnamefont {S.~Y.}\ \bibnamefont {Huang}},
  \bibinfo {author} {\bibfnamefont {J.}~\bibnamefont {Hu}}, \bibinfo {author}
  {\bibfnamefont {R.}~\bibnamefont {Wu}}, \ and\ \bibinfo {author}
  {\bibfnamefont {C.~L.}\ \bibnamefont {Chien}},\ }\href {\doibase
  10.1103/PhysRevLett.110.067206} {\bibfield  {journal} {\bibinfo  {journal}
  {Phys. Rev. Lett.}\ }\textbf {\bibinfo {volume} {110}},\ \bibinfo {pages}
  {067206} (\bibinfo {year} {2013})}\BibitemShut {NoStop}%
\bibitem [{\citenamefont {Rezende}\ \emph {et~al.}(2014)\citenamefont
  {Rezende}, \citenamefont {Rodr\'{i}guez-Su\'{a}rez}, \citenamefont {Cunha},
  \citenamefont {Rodrigues}, \citenamefont {Machado}, \citenamefont
  {Fonseca~Guerra}, \citenamefont {Lopez~Ortiz},\ and\ \citenamefont
  {Azevedo}}]{PhysRevB.89.014416}%
  \BibitemOpen
  \bibfield  {author} {\bibinfo {author} {\bibfnamefont {S.~M.}\ \bibnamefont
  {Rezende}}, \bibinfo {author} {\bibfnamefont {R.~L.}\ \bibnamefont
  {Rodr\'{i}guez-Su\'{a}rez}}, \bibinfo {author} {\bibfnamefont {R.~O.}\
  \bibnamefont {Cunha}}, \bibinfo {author} {\bibfnamefont {A.~R.}\ \bibnamefont
  {Rodrigues}}, \bibinfo {author} {\bibfnamefont {F.~L.~A.}\ \bibnamefont
  {Machado}}, \bibinfo {author} {\bibfnamefont {G.~A.}\ \bibnamefont
  {Fonseca~Guerra}}, \bibinfo {author} {\bibfnamefont {J.~C.}\ \bibnamefont
  {Lopez~Ortiz}}, \ and\ \bibinfo {author} {\bibfnamefont {A.}~\bibnamefont
  {Azevedo}},\ }\href {\doibase 10.1103/PhysRevB.89.014416} {\bibfield
  {journal} {\bibinfo  {journal} {Phys. Rev. B}\ }\textbf {\bibinfo {volume}
  {89}},\ \bibinfo {pages} {014416} (\bibinfo {year} {2014})}\BibitemShut
  {NoStop}%
\bibitem [{\citenamefont {Kajiwara}\ \emph {et~al.}(2010)\citenamefont
  {Kajiwara}, \citenamefont {Harii}, \citenamefont {Takahashi}, \citenamefont
  {Ohe}, \citenamefont {Uchida}, \citenamefont {Mizuguchi}, \citenamefont
  {Umezawa}, \citenamefont {Kawai}, \citenamefont {Ando}, \citenamefont
  {Takanashi}, \citenamefont {Maekawa},\ and\ \citenamefont
  {Saitoh}}]{kajiwaratransmission2010}%
  \BibitemOpen
  \bibfield  {author} {\bibinfo {author} {\bibfnamefont {Y.}~\bibnamefont
  {Kajiwara}}, \bibinfo {author} {\bibfnamefont {K.}~\bibnamefont {Harii}},
  \bibinfo {author} {\bibfnamefont {S.}~\bibnamefont {Takahashi}}, \bibinfo
  {author} {\bibfnamefont {J.}~\bibnamefont {Ohe}}, \bibinfo {author}
  {\bibfnamefont {K.}~\bibnamefont {Uchida}}, \bibinfo {author} {\bibfnamefont
  {M.}~\bibnamefont {Mizuguchi}}, \bibinfo {author} {\bibfnamefont
  {H.}~\bibnamefont {Umezawa}}, \bibinfo {author} {\bibfnamefont
  {H.}~\bibnamefont {Kawai}}, \bibinfo {author} {\bibfnamefont
  {K.}~\bibnamefont {Ando}}, \bibinfo {author} {\bibfnamefont {K.}~\bibnamefont
  {Takanashi}}, \bibinfo {author} {\bibfnamefont {S.}~\bibnamefont {Maekawa}},
  \ and\ \bibinfo {author} {\bibfnamefont {E.}~\bibnamefont {Saitoh}},\ }\href
  {\doibase 10.1038/nature08876} {\bibfield  {journal} {\bibinfo  {journal}
  {Nature}\ }\textbf {\bibinfo {volume} {464}},\ \bibinfo {pages} {262}
  (\bibinfo {year} {2010})}\BibitemShut {NoStop}%
\bibitem [{\citenamefont {Mosendz}\ \emph {et~al.}(2010)\citenamefont
  {Mosendz}, \citenamefont {Pearson}, \citenamefont {Fradin}, \citenamefont
  {Bauer}, \citenamefont {Bader},\ and\ \citenamefont
  {Hoffmann}}]{PhysRevLett.104.046601}%
  \BibitemOpen
  \bibfield  {author} {\bibinfo {author} {\bibfnamefont {O.}~\bibnamefont
  {Mosendz}}, \bibinfo {author} {\bibfnamefont {J.~E.}\ \bibnamefont
  {Pearson}}, \bibinfo {author} {\bibfnamefont {F.~Y.}\ \bibnamefont {Fradin}},
  \bibinfo {author} {\bibfnamefont {G.~E.~W.}\ \bibnamefont {Bauer}}, \bibinfo
  {author} {\bibfnamefont {S.~D.}\ \bibnamefont {Bader}}, \ and\ \bibinfo
  {author} {\bibfnamefont {A.}~\bibnamefont {Hoffmann}},\ }\href {\doibase
  10.1103/PhysRevLett.104.046601} {\bibfield  {journal} {\bibinfo  {journal}
  {Phys. Rev. Lett.}\ }\textbf {\bibinfo {volume} {104}},\ \bibinfo {pages}
  {046601} (\bibinfo {year} {2010})}\BibitemShut {NoStop}%
\bibitem [{\citenamefont {Vilela-Le\~{a}o}\ \emph {et~al.}(2011)\citenamefont
  {Vilela-Le\~{a}o}, \citenamefont {da~Silva}, \citenamefont {Salvador},
  \citenamefont {Rezende},\ and\ \citenamefont
  {Azevedo}}]{:/content/aip/journal/jap/109/7/10.1063/1.3549582}%
  \BibitemOpen
  \bibfield  {author} {\bibinfo {author} {\bibfnamefont {L.~H.}\ \bibnamefont
  {Vilela-Le\~{a}o}}, \bibinfo {author} {\bibfnamefont {G.~L.}\ \bibnamefont
  {da~Silva}}, \bibinfo {author} {\bibfnamefont {C.}~\bibnamefont {Salvador}},
  \bibinfo {author} {\bibfnamefont {S.~M.}\ \bibnamefont {Rezende}}, \ and\
  \bibinfo {author} {\bibfnamefont {A.}~\bibnamefont {Azevedo}},\ }\href@noop
  {} {\bibfield  {journal} {\bibinfo  {journal} {J. Appl. Phys.}\ }\textbf
  {\bibinfo {volume} {109}},\  (\bibinfo {year} {2011})}\BibitemShut {NoStop}%
\bibitem [{\citenamefont {Du}\ \emph {et~al.}(2013)\citenamefont {Du},
  \citenamefont {Wang}, \citenamefont {Pu}, \citenamefont {Meyer},
  \citenamefont {Woodward}, \citenamefont {Yang},\ and\ \citenamefont
  {Hammel}}]{PhysRevLett.111.247202}%
  \BibitemOpen
  \bibfield  {author} {\bibinfo {author} {\bibfnamefont {C.~H.}\ \bibnamefont
  {Du}}, \bibinfo {author} {\bibfnamefont {H.~L.}\ \bibnamefont {Wang}},
  \bibinfo {author} {\bibfnamefont {Y.}~\bibnamefont {Pu}}, \bibinfo {author}
  {\bibfnamefont {T.~L.}\ \bibnamefont {Meyer}}, \bibinfo {author}
  {\bibfnamefont {P.~M.}\ \bibnamefont {Woodward}}, \bibinfo {author}
  {\bibfnamefont {F.~Y.}\ \bibnamefont {Yang}}, \ and\ \bibinfo {author}
  {\bibfnamefont {P.~C.}\ \bibnamefont {Hammel}},\ }\href {\doibase
  10.1103/PhysRevLett.111.247202} {\bibfield  {journal} {\bibinfo  {journal}
  {Phys. Rev. Lett.}\ }\textbf {\bibinfo {volume} {111}},\ \bibinfo {pages}
  {247202} (\bibinfo {year} {2013})}\BibitemShut {NoStop}%
\bibitem [{\citenamefont {Weiler}\ \emph {et~al.}(2014)\citenamefont {Weiler},
  \citenamefont {Shaw}, \citenamefont {Nembach},\ and\ \citenamefont
  {Silva}}]{1401.6469}%
  \BibitemOpen
  \bibfield  {author} {\bibinfo {author} {\bibfnamefont {M.}~\bibnamefont
  {Weiler}}, \bibinfo {author} {\bibfnamefont {J.~M.}\ \bibnamefont {Shaw}},
  \bibinfo {author} {\bibfnamefont {H.~T.}\ \bibnamefont {Nembach}}, \ and\
  \bibinfo {author} {\bibfnamefont {T.~J.}\ \bibnamefont {Silva}},\ }\href@noop
  {} {} (\bibinfo {year} {2014}),\ \Eprint
  {http://arxiv.org/abs/arXiv:1401.6469} {arXiv:1401.6469} \BibitemShut
  {NoStop}%
\bibitem [{\citenamefont {Wei}\ \emph {et~al.}(2013)\citenamefont {Wei},
  \citenamefont {Obstbaum}, \citenamefont {Back},\ and\ \citenamefont
  {Woltersdorf}}]{1307.2961}%
  \BibitemOpen
  \bibfield  {author} {\bibinfo {author} {\bibfnamefont {D.}~\bibnamefont
  {Wei}}, \bibinfo {author} {\bibfnamefont {M.}~\bibnamefont {Obstbaum}},
  \bibinfo {author} {\bibfnamefont {C.}~\bibnamefont {Back}}, \ and\ \bibinfo
  {author} {\bibfnamefont {G.}~\bibnamefont {Woltersdorf}},\ }\href@noop {} {}
  (\bibinfo {year} {2013}),\ \Eprint {http://arxiv.org/abs/arXiv:1307.2961}
  {arXiv:1307.2961} \BibitemShut {NoStop}%
\bibitem [{\citenamefont {Hahn}\ \emph
  {et~al.}(2013{\natexlab{a}})\citenamefont {Hahn}, \citenamefont {de~Loubens},
  \citenamefont {Viret}, \citenamefont {Klein}, \citenamefont {Naletov},\ and\
  \citenamefont {Ben~Youssef}}]{PhysRevLett.111.217204}%
  \BibitemOpen
  \bibfield  {author} {\bibinfo {author} {\bibfnamefont {C.}~\bibnamefont
  {Hahn}}, \bibinfo {author} {\bibfnamefont {G.}~\bibnamefont {de~Loubens}},
  \bibinfo {author} {\bibfnamefont {M.}~\bibnamefont {Viret}}, \bibinfo
  {author} {\bibfnamefont {O.}~\bibnamefont {Klein}}, \bibinfo {author}
  {\bibfnamefont {V.~V.}\ \bibnamefont {Naletov}}, \ and\ \bibinfo {author}
  {\bibfnamefont {J.}~\bibnamefont {Ben~Youssef}},\ }\href {\doibase
  10.1103/PhysRevLett.111.217204} {\bibfield  {journal} {\bibinfo  {journal}
  {Phys. Rev. Lett.}\ }\textbf {\bibinfo {volume} {111}},\ \bibinfo {pages}
  {217204} (\bibinfo {year} {2013}{\natexlab{a}})}\BibitemShut {NoStop}%
\bibitem [{\citenamefont {Vlietstra}\ \emph {et~al.}(2013)\citenamefont
  {Vlietstra}, \citenamefont {Shan}, \citenamefont {Castel}, \citenamefont {van
  Wees},\ and\ \citenamefont {Ben~Youssef}}]{PhysRevB.87.184421}%
  \BibitemOpen
  \bibfield  {author} {\bibinfo {author} {\bibfnamefont {N.}~\bibnamefont
  {Vlietstra}}, \bibinfo {author} {\bibfnamefont {J.}~\bibnamefont {Shan}},
  \bibinfo {author} {\bibfnamefont {V.}~\bibnamefont {Castel}}, \bibinfo
  {author} {\bibfnamefont {B.~J.}\ \bibnamefont {van Wees}}, \ and\ \bibinfo
  {author} {\bibfnamefont {J.}~\bibnamefont {Ben~Youssef}},\ }\href {\doibase
  10.1103/PhysRevB.87.184421} {\bibfield  {journal} {\bibinfo  {journal} {Phys.
  Rev. B}\ }\textbf {\bibinfo {volume} {87}},\ \bibinfo {pages} {184421}
  (\bibinfo {year} {2013})}\BibitemShut {NoStop}%
\bibitem [{\citenamefont {Hahn}\ \emph
  {et~al.}(2013{\natexlab{b}})\citenamefont {Hahn}, \citenamefont {de~Loubens},
  \citenamefont {Klein}, \citenamefont {Viret}, \citenamefont {Naletov},\ and\
  \citenamefont {Ben~Youssef}}]{PhysRevB.87.174417}%
  \BibitemOpen
  \bibfield  {author} {\bibinfo {author} {\bibfnamefont {C.}~\bibnamefont
  {Hahn}}, \bibinfo {author} {\bibfnamefont {G.}~\bibnamefont {de~Loubens}},
  \bibinfo {author} {\bibfnamefont {O.}~\bibnamefont {Klein}}, \bibinfo
  {author} {\bibfnamefont {M.}~\bibnamefont {Viret}}, \bibinfo {author}
  {\bibfnamefont {V.~V.}\ \bibnamefont {Naletov}}, \ and\ \bibinfo {author}
  {\bibfnamefont {J.}~\bibnamefont {Ben~Youssef}},\ }\href {\doibase
  10.1103/PhysRevB.87.174417} {\bibfield  {journal} {\bibinfo  {journal} {Phys.
  Rev. B}\ }\textbf {\bibinfo {volume} {87}},\ \bibinfo {pages} {174417}
  (\bibinfo {year} {2013}{\natexlab{b}})}\BibitemShut {NoStop}%
\bibitem [{\citenamefont {Nakayama}\ \emph {et~al.}(2013)\citenamefont
  {Nakayama}, \citenamefont {Althammer}, \citenamefont {Chen}, \citenamefont
  {Uchida}, \citenamefont {Kajiwara}, \citenamefont {Kikuchi}, \citenamefont
  {Ohtani}, \citenamefont {Gepr\"ags}, \citenamefont {Opel}, \citenamefont
  {Takahashi}, \citenamefont {Gross}, \citenamefont {Bauer}, \citenamefont
  {Goennenwein},\ and\ \citenamefont {Saitoh}}]{PhysRevLett.110.206601}%
  \BibitemOpen
  \bibfield  {author} {\bibinfo {author} {\bibfnamefont {H.}~\bibnamefont
  {Nakayama}}, \bibinfo {author} {\bibfnamefont {M.}~\bibnamefont {Althammer}},
  \bibinfo {author} {\bibfnamefont {Y.-T.}\ \bibnamefont {Chen}}, \bibinfo
  {author} {\bibfnamefont {K.}~\bibnamefont {Uchida}}, \bibinfo {author}
  {\bibfnamefont {Y.}~\bibnamefont {Kajiwara}}, \bibinfo {author}
  {\bibfnamefont {D.}~\bibnamefont {Kikuchi}}, \bibinfo {author} {\bibfnamefont
  {T.}~\bibnamefont {Ohtani}}, \bibinfo {author} {\bibfnamefont
  {S.}~\bibnamefont {Gepr\"ags}}, \bibinfo {author} {\bibfnamefont
  {M.}~\bibnamefont {Opel}}, \bibinfo {author} {\bibfnamefont {S.}~\bibnamefont
  {Takahashi}}, \bibinfo {author} {\bibfnamefont {R.}~\bibnamefont {Gross}},
  \bibinfo {author} {\bibfnamefont {G.~E.~W.}\ \bibnamefont {Bauer}}, \bibinfo
  {author} {\bibfnamefont {S.~T.~B.}\ \bibnamefont {Goennenwein}}, \ and\
  \bibinfo {author} {\bibfnamefont {E.}~\bibnamefont {Saitoh}},\ }\href
  {\doibase 10.1103/PhysRevLett.110.206601} {\bibfield  {journal} {\bibinfo
  {journal} {Phys. Rev. Lett.}\ }\textbf {\bibinfo {volume} {110}},\ \bibinfo
  {pages} {206601} (\bibinfo {year} {2013})}\BibitemShut {NoStop}%
\bibitem [{\citenamefont {Althammer}\ \emph {et~al.}(2013)\citenamefont
  {Althammer}, \citenamefont {Meyer}, \citenamefont {Nakayama}, \citenamefont
  {Schreier}, \citenamefont {Altmannshofer}, \citenamefont {Weiler},
  \citenamefont {Huebl}, \citenamefont {Gepr\"ags}, \citenamefont {Opel},
  \citenamefont {Gross}, \citenamefont {Meier}, \citenamefont {Klewe},
  \citenamefont {Kuschel}, \citenamefont {Schmalhorst}, \citenamefont {Reiss},
  \citenamefont {Shen}, \citenamefont {Gupta}, \citenamefont {Chen},
  \citenamefont {Bauer}, \citenamefont {Saitoh},\ and\ \citenamefont
  {Goennenwein}}]{PhysRevB.87.224401}%
  \BibitemOpen
  \bibfield  {author} {\bibinfo {author} {\bibfnamefont {M.}~\bibnamefont
  {Althammer}}, \bibinfo {author} {\bibfnamefont {S.}~\bibnamefont {Meyer}},
  \bibinfo {author} {\bibfnamefont {H.}~\bibnamefont {Nakayama}}, \bibinfo
  {author} {\bibfnamefont {M.}~\bibnamefont {Schreier}}, \bibinfo {author}
  {\bibfnamefont {S.}~\bibnamefont {Altmannshofer}}, \bibinfo {author}
  {\bibfnamefont {M.}~\bibnamefont {Weiler}}, \bibinfo {author} {\bibfnamefont
  {H.}~\bibnamefont {Huebl}}, \bibinfo {author} {\bibfnamefont
  {S.}~\bibnamefont {Gepr\"ags}}, \bibinfo {author} {\bibfnamefont
  {M.}~\bibnamefont {Opel}}, \bibinfo {author} {\bibfnamefont {R.}~\bibnamefont
  {Gross}}, \bibinfo {author} {\bibfnamefont {D.}~\bibnamefont {Meier}},
  \bibinfo {author} {\bibfnamefont {C.}~\bibnamefont {Klewe}}, \bibinfo
  {author} {\bibfnamefont {T.}~\bibnamefont {Kuschel}}, \bibinfo {author}
  {\bibfnamefont {J.-M.}\ \bibnamefont {Schmalhorst}}, \bibinfo {author}
  {\bibfnamefont {G.}~\bibnamefont {Reiss}}, \bibinfo {author} {\bibfnamefont
  {L.}~\bibnamefont {Shen}}, \bibinfo {author} {\bibfnamefont {A.}~\bibnamefont
  {Gupta}}, \bibinfo {author} {\bibfnamefont {Y.-T.}\ \bibnamefont {Chen}},
  \bibinfo {author} {\bibfnamefont {G.~E.~W.}\ \bibnamefont {Bauer}}, \bibinfo
  {author} {\bibfnamefont {E.}~\bibnamefont {Saitoh}}, \ and\ \bibinfo {author}
  {\bibfnamefont {S.~T.~B.}\ \bibnamefont {Goennenwein}},\ }\href {\doibase
  10.1103/PhysRevB.87.224401} {\bibfield  {journal} {\bibinfo  {journal} {Phys.
  Rev. B}\ }\textbf {\bibinfo {volume} {87}},\ \bibinfo {pages} {224401}
  (\bibinfo {year} {2013})}\BibitemShut {NoStop}%
\bibitem [{\citenamefont {Chen}\ \emph {et~al.}(2013)\citenamefont {Chen},
  \citenamefont {Takahashi}, \citenamefont {Nakayama}, \citenamefont
  {Althammer}, \citenamefont {Goennenwein}, \citenamefont {Saitoh},\ and\
  \citenamefont {Bauer}}]{PhysRevB.87.144411}%
  \BibitemOpen
  \bibfield  {author} {\bibinfo {author} {\bibfnamefont {Y.-T.}\ \bibnamefont
  {Chen}}, \bibinfo {author} {\bibfnamefont {S.}~\bibnamefont {Takahashi}},
  \bibinfo {author} {\bibfnamefont {H.}~\bibnamefont {Nakayama}}, \bibinfo
  {author} {\bibfnamefont {M.}~\bibnamefont {Althammer}}, \bibinfo {author}
  {\bibfnamefont {S.~T.~B.}\ \bibnamefont {Goennenwein}}, \bibinfo {author}
  {\bibfnamefont {E.}~\bibnamefont {Saitoh}}, \ and\ \bibinfo {author}
  {\bibfnamefont {G.~E.~W.}\ \bibnamefont {Bauer}},\ }\href {\doibase
  10.1103/PhysRevB.87.144411} {\bibfield  {journal} {\bibinfo  {journal} {Phys.
  Rev. B}\ }\textbf {\bibinfo {volume} {87}},\ \bibinfo {pages} {144411}
  (\bibinfo {year} {2013})}\BibitemShut {NoStop}%
\bibitem [{\citenamefont {Kampfrath}\ \emph {et~al.}(2013)\citenamefont
  {Kampfrath}, \citenamefont {Battiato}, \citenamefont {Maldonado},
  \citenamefont {Eilers}, \citenamefont {N\"otzold}, \citenamefont
  {M\"ahrlein}, \citenamefont {Zbarsky}, \citenamefont {Freimuth},
  \citenamefont {Mokrousov}, \citenamefont {Bl\"ugel}, \citenamefont {Wolf},
  \citenamefont {Radu}, \citenamefont {Oppeneer},\ and\ \citenamefont
  {M\"unzenberg}}]{kampfrathterahertz2013}%
  \BibitemOpen
  \bibfield  {author} {\bibinfo {author} {\bibfnamefont {T.}~\bibnamefont
  {Kampfrath}}, \bibinfo {author} {\bibfnamefont {M.}~\bibnamefont {Battiato}},
  \bibinfo {author} {\bibfnamefont {P.}~\bibnamefont {Maldonado}}, \bibinfo
  {author} {\bibfnamefont {G.}~\bibnamefont {Eilers}}, \bibinfo {author}
  {\bibfnamefont {J.}~\bibnamefont {N\"otzold}}, \bibinfo {author}
  {\bibfnamefont {S.}~\bibnamefont {M\"ahrlein}}, \bibinfo {author}
  {\bibfnamefont {V.}~\bibnamefont {Zbarsky}}, \bibinfo {author} {\bibfnamefont
  {F.}~\bibnamefont {Freimuth}}, \bibinfo {author} {\bibfnamefont
  {Y.}~\bibnamefont {Mokrousov}}, \bibinfo {author} {\bibfnamefont
  {S.}~\bibnamefont {Bl\"ugel}}, \bibinfo {author} {\bibfnamefont
  {M.}~\bibnamefont {Wolf}}, \bibinfo {author} {\bibfnamefont {I.}~\bibnamefont
  {Radu}}, \bibinfo {author} {\bibfnamefont {P.~M.}\ \bibnamefont {Oppeneer}},
  \ and\ \bibinfo {author} {\bibfnamefont {M.}~\bibnamefont {M\"unzenberg}},\
  }\href {\doibase 10.1038/nnano.2013.43} {\bibfield  {journal} {\bibinfo
  {journal} {Nat. Nanotechnol.}\ }\textbf {\bibinfo {volume} {8}},\ \bibinfo
  {pages} {256} (\bibinfo {year} {2013})}\BibitemShut {NoStop}%
\bibitem [{Note1()}]{Note1}%
  \BibitemOpen
  \bibinfo {note} {For the sake of simplicity, we use the term
  ``ferromagnetic'' in the sense of ``exhibiting long-range magnetic order''.
  In particular, we also refer to ferrimagnetic materials as
  ferromagnetic}\BibitemShut {NoStop}%
\bibitem [{\citenamefont {Opel}(2012)}]{0022-3727-45-3-033001}%
  \BibitemOpen
  \bibfield  {author} {\bibinfo {author} {\bibfnamefont {M.}~\bibnamefont
  {Opel}},\ }\href {http://stacks.iop.org/0022-3727/45/i=3/a=033001} {\bibfield
   {journal} {\bibinfo  {journal} {J. Phys. D: Appl. Phys.}\ }\textbf {\bibinfo
  {volume} {45}},\ \bibinfo {pages} {033001} (\bibinfo {year}
  {2012})}\BibitemShut {NoStop}%
\bibitem [{\citenamefont {Pozar}(2011)}]{pozar2011microwave}%
  \BibitemOpen
  \bibfield  {author} {\bibinfo {author} {\bibfnamefont {D.}~\bibnamefont
  {Pozar}},\ }\href {http://books.google.de/books?id=JegbAAAAQBAJ} {\emph
  {\bibinfo {title} {Microwave Engineering, 4th Edition}}}\ (\bibinfo
  {publisher} {Wiley Global Education},\ \bibinfo {year} {2011})\BibitemShut
  {NoStop}%
\bibitem [{\citenamefont {Dunsmore}(2012)}]{dunsmore2012handbook}%
  \BibitemOpen
  \bibfield  {author} {\bibinfo {author} {\bibfnamefont {J.}~\bibnamefont
  {Dunsmore}},\ }\href {http://books.google.de/books?id=GdDrTH8YE1kC} {\emph
  {\bibinfo {title} {Handbook of Microwave Component Measurements: with
  Advanced VNA Techniques}}}\ (\bibinfo  {publisher} {Wiley},\ \bibinfo {year}
  {2012})\BibitemShut {NoStop}%
\bibitem [{\citenamefont {Sengupta}\ and\ \citenamefont
  {Liepa}(2005)}]{sengupta2005applied}%
  \BibitemOpen
  \bibfield  {author} {\bibinfo {author} {\bibfnamefont {D.}~\bibnamefont
  {Sengupta}}\ and\ \bibinfo {author} {\bibfnamefont {V.}~\bibnamefont
  {Liepa}},\ }\href {http://books.google.de/books?id=3spAMbigGIMC} {\emph
  {\bibinfo {title} {Applied Electromagnetics and Electromagnetic
  Compatibility}}},\ Wiley Series in Microwave and Optical Engineering\
  (\bibinfo  {publisher} {Wiley},\ \bibinfo {year} {2005})\BibitemShut
  {NoStop}%
\bibitem [{\citenamefont {Weiler}\ \emph {et~al.}(2013)\citenamefont {Weiler},
  \citenamefont {Althammer}, \citenamefont {Schreier}, \citenamefont {Lotze},
  \citenamefont {Pernpeintner}, \citenamefont {Meyer}, \citenamefont {Huebl},
  \citenamefont {Gross}, \citenamefont {Kamra}, \citenamefont {Xiao},
  \citenamefont {Chen}, \citenamefont {Jiao}, \citenamefont {Bauer},\ and\
  \citenamefont {Goennenwein}}]{PhysRevLett.111.176601}%
  \BibitemOpen
  \bibfield  {author} {\bibinfo {author} {\bibfnamefont {M.}~\bibnamefont
  {Weiler}}, \bibinfo {author} {\bibfnamefont {M.}~\bibnamefont {Althammer}},
  \bibinfo {author} {\bibfnamefont {M.}~\bibnamefont {Schreier}}, \bibinfo
  {author} {\bibfnamefont {J.}~\bibnamefont {Lotze}}, \bibinfo {author}
  {\bibfnamefont {M.}~\bibnamefont {Pernpeintner}}, \bibinfo {author}
  {\bibfnamefont {S.}~\bibnamefont {Meyer}}, \bibinfo {author} {\bibfnamefont
  {H.}~\bibnamefont {Huebl}}, \bibinfo {author} {\bibfnamefont
  {R.}~\bibnamefont {Gross}}, \bibinfo {author} {\bibfnamefont
  {A.}~\bibnamefont {Kamra}}, \bibinfo {author} {\bibfnamefont
  {J.}~\bibnamefont {Xiao}}, \bibinfo {author} {\bibfnamefont {Y.-T.}\
  \bibnamefont {Chen}}, \bibinfo {author} {\bibfnamefont {H.J.}~\bibnamefont
  {Jiao}}, \bibinfo {author} {\bibfnamefont {G.~E.~W.}\ \bibnamefont {Bauer}},
  \ and\ \bibinfo {author} {\bibfnamefont {S.~T.~B.}\ \bibnamefont
  {Goennenwein}},\ }\href {\doibase 10.1103/PhysRevLett.111.176601} {\bibfield
  {journal} {\bibinfo  {journal} {Phys. Rev. Lett.}\ }\textbf {\bibinfo
  {volume} {111}},\ \bibinfo {pages} {176601} (\bibinfo {year}
  {2013})}\BibitemShut {NoStop}%
\bibitem [{\citenamefont {Haney}\ \emph
  {et~al.}(2013{\natexlab{a}})\citenamefont {Haney}, \citenamefont {Lee},
  \citenamefont {Lee}, \citenamefont {Manchon},\ and\ \citenamefont
  {Stiles}}]{PhysRevB.87.174411}%
  \BibitemOpen
  \bibfield  {author} {\bibinfo {author} {\bibfnamefont {P.~M.}\ \bibnamefont
  {Haney}}, \bibinfo {author} {\bibfnamefont {H.-W.}\ \bibnamefont {Lee}},
  \bibinfo {author} {\bibfnamefont {K.-J.}\ \bibnamefont {Lee}}, \bibinfo
  {author} {\bibfnamefont {A.}~\bibnamefont {Manchon}}, \ and\ \bibinfo
  {author} {\bibfnamefont {M.~D.}\ \bibnamefont {Stiles}},\ }\href {\doibase
  10.1103/PhysRevB.87.174411} {\bibfield  {journal} {\bibinfo  {journal} {Phys.
  Rev. B}\ }\textbf {\bibinfo {volume} {87}},\ \bibinfo {pages} {174411}
  (\bibinfo {year} {2013}{\natexlab{a}})}\BibitemShut {NoStop}%
\bibitem [{\citenamefont {Haney}\ \emph
  {et~al.}(2013{\natexlab{b}})\citenamefont {Haney}, \citenamefont {Lee},
  \citenamefont {Lee}, \citenamefont {Manchon},\ and\ \citenamefont
  {Stiles}}]{PhysRevB.88.214417}%
  \BibitemOpen
  \bibfield  {author} {\bibinfo {author} {\bibfnamefont {P.~M.}\ \bibnamefont
  {Haney}}, \bibinfo {author} {\bibfnamefont {H.-W.}\ \bibnamefont {Lee}},
  \bibinfo {author} {\bibfnamefont {K.-J.}\ \bibnamefont {Lee}}, \bibinfo
  {author} {\bibfnamefont {A.}~\bibnamefont {Manchon}}, \ and\ \bibinfo
  {author} {\bibfnamefont {M.~D.}\ \bibnamefont {Stiles}},\ }\href {\doibase
  10.1103/PhysRevB.88.214417} {\bibfield  {journal} {\bibinfo  {journal} {Phys.
  Rev. B}\ }\textbf {\bibinfo {volume} {88}},\ \bibinfo {pages} {214417}
  (\bibinfo {year} {2013}{\natexlab{b}})}\BibitemShut {NoStop}%
\bibitem [{\citenamefont {Sakurai}(1994)}]{sakurai1994modern}%
  \BibitemOpen
  \bibfield  {author} {\bibinfo {author} {\bibfnamefont {J.}~\bibnamefont
  {Sakurai}},\ }\href@noop {} {\emph {\bibinfo {title} {Modern Quantum
  Mechanics}}}\ (\bibinfo  {publisher} {Addison-Wesley},\ \bibinfo {year}
  {1994})\BibitemShut {NoStop}%
\bibitem [{\citenamefont {Takahashi}\ and\ \citenamefont
  {Maekawa}(2008)}]{1468-6996-9-1-014105}%
  \BibitemOpen
  \bibfield  {author} {\bibinfo {author} {\bibfnamefont {S.}~\bibnamefont
  {Takahashi}}\ and\ \bibinfo {author} {\bibfnamefont {S.}~\bibnamefont
  {Maekawa}},\ }\href {http://stacks.iop.org/1468-6996/9/i=1/a=014105}
  {\bibfield  {journal} {\bibinfo  {journal} {Sci. Tech. Adv. Mater.}\ }\textbf
  {\bibinfo {volume} {9}},\ \bibinfo {pages} {014105} (\bibinfo {year}
  {2008})}\BibitemShut {NoStop}%
\end{thebibliography}
\end{document}